\documentclass[12pt,preprint]{aastex}

\newcommand\psr{PSR~J1718$-$3718}
\def\lapp{\ifmmode\stackrel{<}{_{\sim}}\else$\stackrel{<}{_{\sim}}$\fi}
\def\gapp{\ifmmode\stackrel{>}{_{\sim}}\else$\stackrel{>}{_{\sim}}$\fi}

\begin{document}

\title{{\it Chandra} X-ray Detection of the High-Magnetic-Field Radio Pulsar
PSR~J1718$-$3718}

\author{
V. M. Kaspi\altaffilmark{1,2} \&
M. A. McLaughlin\altaffilmark{3}
}

\altaffiltext{1}{Department of Physics, Rutherford Physics Building,
McGill University, 3600 University Street, Montreal, Quebec,
H3A 2T8, Canada}

\altaffiltext{2}{Canada Research Chair; NSERC Steacie Fellow}

\altaffiltext{3}{Jodrell Bank Observatory, University of Manchester, Macclesfield, Cheshire, SK11 9DL, UK}

\begin{abstract}
We report on the serendipitous X-ray detection, using the {\it Chandra
X-ray Observatory,} of the radio pulsar \psr.
This pulsar has one of the highest inferred surface dipole magnetic fields
in the radio pulsar population ($B = 7.4 \times 10^{13}$~G), higher than
that inferred for one well-known Anomalous X-ray Pulsar (AXP).  The X-ray
emission for \psr\ appears point-like and has a purely thermal spectrum, with $kT =
0.145^{+0.053}_{-0.020}$~keV and absorbed 0.5--2~keV flux of $(6.3-6.9)
\times 10^{-15}$~erg~s$^{-1}$~cm$^{-2}$.  We show that the pulsar's
2--10~keV luminosity is several orders of magnitude smaller than those of the
non-transient AXPs, and consistent with the predictions of standard
models for initial cooling. The number of high-magnetic-field
radio pulsars observed at X-ray energies now stands at five. All are X-ray
faint, suggesting that either there is a significant physical distinction
between high-magnetic-field radio pulsars and AXPs, or that
high-magnetic-field radio pulsars are, in fact, quiescent AXPs.
\end{abstract}

\keywords{pulsars: general --- pulsars: individual (PSR J1718$-$3718) --- stars: neutron}

\section{Introduction}

The existence of magnetars -- young, isolated, high-magnetic-field neutron
stars -- is now well supported by a variety of independent lines of
evidence.  For recent reviews, see \citet{wt04} or \citet{kg04a}. There
appear to be at least two flavors of magnetar:  soft-gamma repeaters
(SGRs) and anomalous X-ray pulsars (AXPs).  Defining properties of both
are their X-ray pulsations having luminosity in the range $10^{34} -
10^{36}$~erg~s$^{-1}$, periods ranging from 6 -- 12~s, period
derivatives of $10^{-13}-10^{-11}$, and surface dipolar magnetic fields in
the range $0.6 - 7 \times 10^{14}$~G, assuming the vacuum dipole model
formula for magnetic braking\footnote{Throughout the paper, magnetic
fields discussed are calculated via $B\equiv 3.2 \times 10^{19} \sqrt{P
\dot{P}}$~G, where $P$ is the spin period and $\dot{P}$ the period
derivative}. In the magnetar model, the pulsed X-rays are likely the
combined result of surface thermal emission \citep[e.g.][]{oze03,lh03a}, with a
non-thermal high-energy tail resulting from resonant scattering of thermal
photons off magnetospheric currents \citep{tlk02}.  The X-rays, in the
magnetar model, are ultimately powered by an internally decaying very
strong magnetic field. Despite numerous attempts, no magnetars have been
detected at radio frequencies \citep{kbhc85,cjl94,llc98,gsg01}, which has been suggested
as implying that pair production ceases above some critical magnetic field
\citep{zh00}.

An open issue in the magnetar model is the connection of these X-ray
sources to radio pulsars. One might expect high-$B$ radio pulsars to be
more X-ray bright than low-$B$ sources, and possibly exhibit magnetar-like
X-ray emission. \citet{pkc00} searched for enhanced X-ray emission from
the high-$B$ ($5.5 \times 10^{13}$~G) radio pulsar PSR~J1814$-$1744, and
placed an upper limit on its X-ray luminosity that was much lower than
those of the five then-known AXPs (4U 0142+61, 1E 1048$-$9537, RXS
1708$-$4009, 1E 1841$-$045, 1E 2259+586).  \citet{gklp04} showed that the
nearby radio pulsar PSR~B0154+61 ($B=2.1 \times 10^{13}$~G) has an X-ray
luminosity 2--3 orders of magnitude lower than those of the same five
AXPs.  \citet{msk+03} reported on X-ray observations of PSR~J1847$-$0130
($B = 9.4 \times 10^{13}$~G), which has the highest inferred surface
dipolar magnetic field of any known radio pulsar, and calculated an upper limit
on its X-ray luminosity that was lower than those of all but
one of the above five AXPs.  \citet{gs03} studied PSR~J1119$-$6127 ($B = 4.4
\times 10^{13}$~G), also finding it to be X-ray underluminous relative to
the standard AXP group.

There are several possible ways to explain these results. 
There could exist a well-defined critical $B$ field above which the
magnetar mechanism abruptly turns on.  However, that would also require
that $B$ fields inferred from spin-down are unreliable at the factor of
$\gapp$2 level, given the overlap in high-$B$ radio pulsar fields and those
of the AXPs (e.g. 1E~2259+586 has $B=6 \times 10^{13}$~G).  It could also
be that AXPs and SGRs have higher-order multipole moments that go
undetected in spin-down, such that their true surface fields are orders of
magnitude higher.  The recently revealed strong X-ray variability seen in
some AXPs \citep[e.g.][]{ims+04,gk04} suggests that magnetar emission
could be transient in many high-$B$ neutron stars.  Of course, which
neutron stars become magnetars could depend on other, currently ``hidden''
neutron-star properties besides $B$ field, such as mass.

\psr\ is a radio pulsar that was recently discovered in the Parkes
Multibeam Survey \citep{hfs+04}.  It has spin period $P=3.3$~s and a
spin-down rate of $\dot{P}=1.5\times 10^{-12}$, which imply a
characteristic age $\tau_c \equiv P/2\dot{P} = 34$~kyr, spin-down
luminosity $\dot{E}\equiv 4\pi^2 I \dot{P}/P^3 = 1.6 \times
10^{33}$~erg~s$^{-1}$, and a surface dipolar magnetic field of $7.4 \times
10^{13}$~G. Its inferred magnetic field is the second highest of all known
radio pulsars and is higher than that of the well established AXP
1E~2259+586. Here we report on the first X-ray detection of this pulsar in
a deep {\it Chandra X-ray Observatory} observation of a nearby field.

\section{Observations and Results}

The position of \psr\ was observed serendipitously by {\it Chandra} in an
ACIS-S Timed Exposure (TE) obtained on 2002 May 13.  The observation (PI
P. Slane, Sequence Number 500235) had as its target the unrelated
supernova remnant G347.7+0.2.  The nominal telescope pointing was $7.0'$
away from the pulsar's position derived from radio timing.  As a result,
the position of \psr\ lies on Chip 6, far from the optical axis, where the
mirror point-spread-function (PSF) is significantly extended and distorted
asymmetrically.

We obtained the public data set using the {\it Chandra} Science Center's
{\it WebChaser} facility, and reduced the data with the CIAO software
package (version 3.1), with calibration database CALDB version 2.28. After
standard filtering using CIAO threads for ACIS-S
data\footnote{http://asc.harvard.edu/ciao/threads/index.html}, the
effective integration time was 55.7~ks.

\subsection{Imaging}

\begin{figure}
\plotone{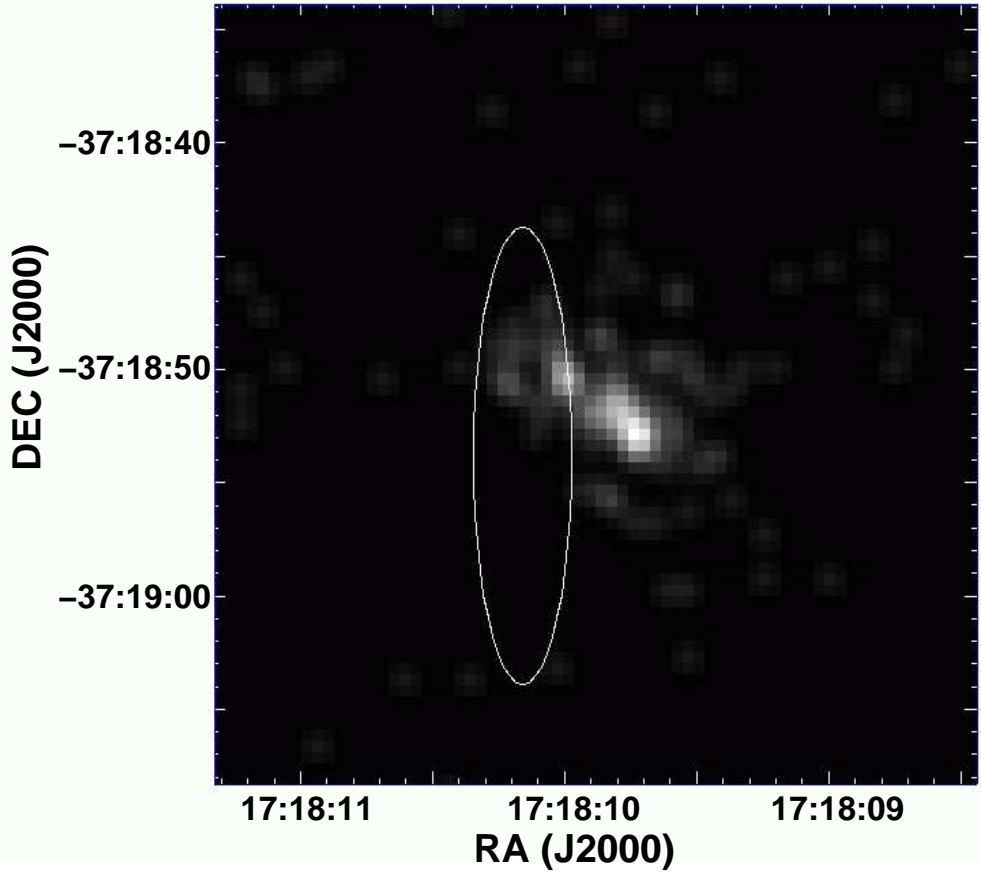}
\figcaption{{\it Chandra} image of the \psr\ field in the 0.5--3.0~keV band.
The image has been smoothed with a Gaussian kernel having $\sigma = 1$~pixel.
Although the source appears extended, its size and morphology, including
angle of asymmetry, are consistent
with the 1.5~keV PSF at this detector position.  The formal {\tt TEMPO} 2$\sigma$ error 
region of the radio
timing position is shown with an ellipse.  See text for details.
\label{fig:image}
}
\end{figure}

The X-ray emission as seen by {\it Chandra} around the radio position
of \psr\ is shown in Figure~\ref{fig:image}.  The source is identified
with CIAO's {\tt celldetect} routine as having a signal-to-noise ratio
of 6.4, for events in the energy range 0.5--3.0~keV.  No source is
apparent in images made with events having energies $>3.0$~keV.
Although the source appears extended (Fig.~\ref{fig:image}), given its
large off-axis angle, its extent both in size and morphology, including
the angle of asymmetry, is consistent with the instrumental PSF at
1.5~keV, as determined using the CIAO {\tt mkpsf} routine.  Indeed,
using counts in the range 0.5--3.0~keV, {\tt celldetect} run with
default parameters reports a ratio of source to PSF size of 1.01.
Given that the approximate 95\% encircled energy radius for an object
$7.0'$ off axis is $\sim 7''$, we cannot rule out the presence of faint
emission having extent significantly smaller than this.  However, as
argued below, the spectrum strongly favors the emission originating
from a point source.

The {\tt celldetect} routine reports a best-fit position for the X-ray
source of (J2000) RA = 17$^{\rm h}$18$^{\rm m}$9$^{\rm s}$.84$\pm$0.02,
DEC = $-37^{\circ}$18$'$51$''$.6$\pm$0.2.  These (1$\sigma$) uncertainties are
statistical, and do not include the systematic uncertainty in {\it
Chandra}'s pointing.  Note that for sources that are within $3'$ of the
aimpoint, the 90\% uncertainty circle of {\it Chandra}'s absolute pointing
has radius\footnote{http://cxc.harvard.edu/cal/ASPECT/celmon/} 0.6$''$.
For sources, like ours, that are further off-axis, the absolute pointing
uncertainty has not been well determined.  This is an important caveat.

A timing analysis of the radio data \citep[see][for a description of the
data and its analysis]{hfs+04} yields a radio timing position of (J2000)
RA = 17$^{\rm h}$18$^{\rm m}$10$^{\rm s}$.162$\pm$0.194, DEC =
$-37^{\circ}$18$'$53$''$.75$\pm$10.0, where the quoted errors are formal
$2\sigma$ uncertainties as reported by {\tt TEMPO}, and 10 months of
additional timing data have been included since the most recently
published result.  Doubling the formal {\tt TEMPO} uncertainties when
reporting timing parameter errors is standard practice and is done to account
for likely contamination from timing noise.  Indeed, like most young
pulsars, \psr\ exhibits significant timing noise (RMS 74~ms after fitting
for position, $P$ and $\dot{P}$), so the above-quoted
uncertainties are likely to be good approximations to the true $1\sigma$
uncertainties. The formal positional offset in declination is therefore
$2.2''$, or $\sim 0.2\sigma$, while the RA offset is 0$^{\rm
s}$.32, or $\sim 1.6 \sigma$.  Note that these numbers do {\it not}
include the unknown {\it Chandra} pointing uncertainty so are lower limits
only.  We conclude that the source positions are consistent within the
uncertainties.

However, given the slight possible positional offset, as well as the
absence of unambiguous proof of the association via the detection of X-ray
pulsations at the radio period (not possible with the ACIS-S data because
it has effective time resolution of 3.2~s), it is reasonable to question
if the X-ray source is really associated with the radio pulsar.  We can
estimate the probability of chance superposition using a log~N/log~S
relationship for {\it Chandra} sources in the 0.5--2.0~keV band,
appropriate for this source \citep{glv+03}.  In this relation, flux is the
unabsorbed value; thus the probability of an X-ray source being near the
pulsar position purely by chance depends strongly on the former's spectral
parameters. As we show below, given only 110 source counts, these
parameters are not well determined.  However, even for the lowest
plausible unabsorbed source flux for our source, the log~N/log~S relation
predicts $\sim$180 sources per square degree.  With timing noise so strong
in this pulsar, we would likely consider positional agreement within
$\sim$10$''$ to be a plausible association.  In this case, the probability
of a random source in this area of sky is only 1\%.  That the offset is
smaller than 10$''$, as well as that the unabsorbed flux is likely
significantly larger than the lowest reasonable value (see below) make
this 1\% probability likely to be a large overestimate.  Thus, the
association appears extremely likely.  We further note that the nearest
optical counterpart in the uncalibrated plates of the Sloan Digital Sky
Survey \citep{pmh+03}, with a limiting magnitude of $\sim$ 22, is more
than 20$''$ away, well outside of our {\it Chandra} error radius.

\subsection{Spectroscopy}

Counts from the pulsar were extracted using an elliptical extraction region
having semi-major and semi-minor axes of 26 and 18 pixels (13$''$ and 9$''$), respectively,
rotated to angle 308$^{\circ}$ west of north.  A nearby, non-overlapping
source-free region having the same elliptical shape and orientation, but
with semi-major and semi-minor axes 40 and 32 pixels (20$''$ and 16$''$), respectively, was
used to estimate the background.  The total number of source counts after
background subtraction was 110, implying a count rate of
0.00197$\pm$0.00019~cps.

\begin{figure}
\plotone{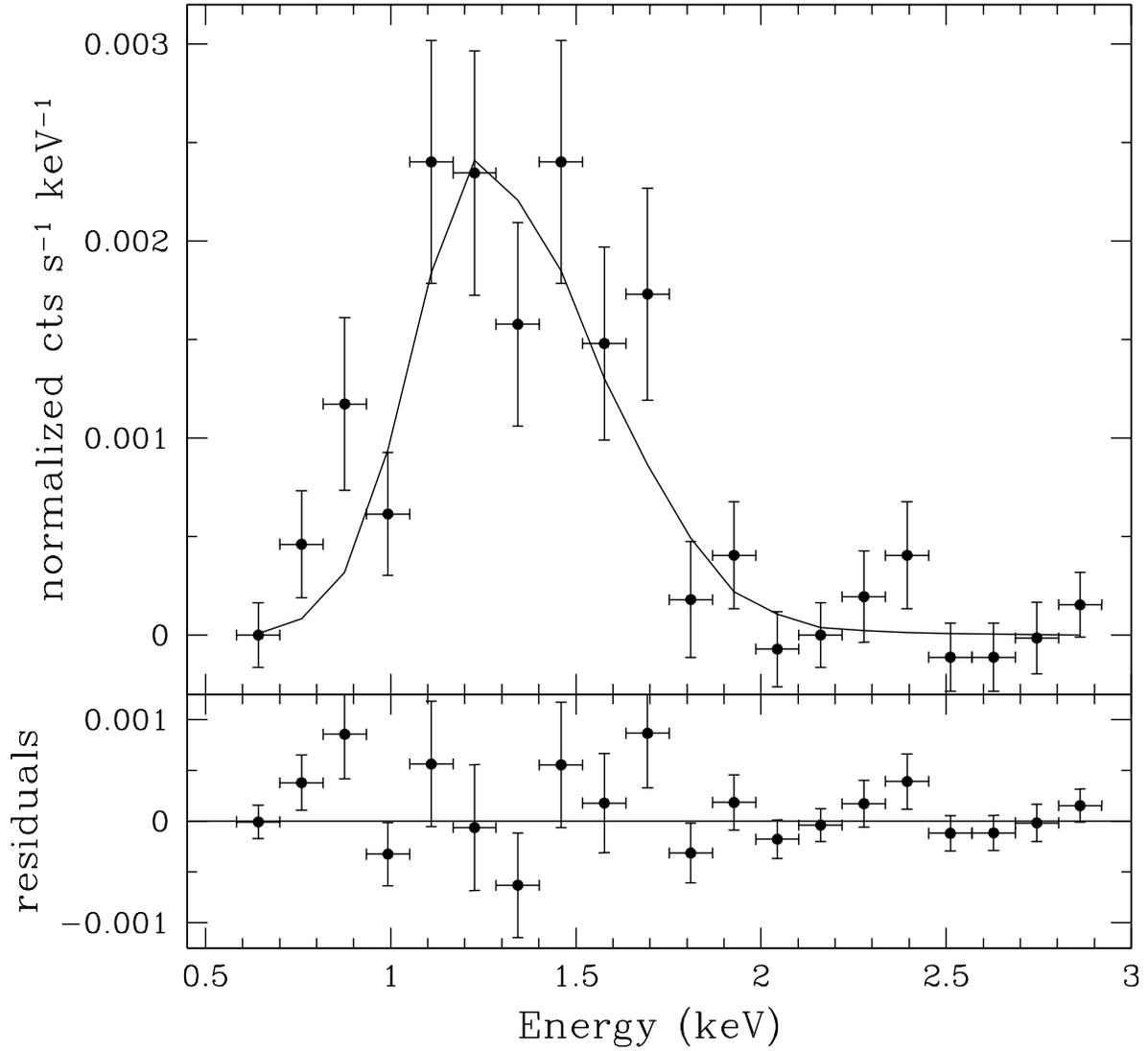}
\figcaption{
Top panel: Spectrum of the X-ray counterpart to \psr, with best-fit model
plotted with a solid line (see Table \protect\ref{ta:spectrum} for best-fit parameters).
Bottom panel: Residuals from the best-fit model.
\label{fig:spectrum}
}
\end{figure}

RMF and ARF files were generated for the source and background using the
CIAO script {\tt psextract}, and spectra grouped by a factor of 8 were fed
into the spectral fitting package {\it XSPEC} (version 11.3.1).  Spectral
channels having energies below 0.5~keV and above 3.0~keV were ignored.
The data were well described by an absorbed black-body model; best-fit model
parameters are given in Table~\ref{ta:spectrum}, and the spectrum and
best-fit model with residuals are shown in Figure~\ref{fig:spectrum}.
Although a power-law model yielded a statistically acceptable fit, the
best-fit power-law index was $\sim$8--9, rendering such a model
implausible.  This is consistent with the absence of counts above
$\sim$2~keV. Fitting for multi-component models was unreasonable due to
the small number of counts available. However, it is clear that it is clear that the emission 
is dominantly thermal in origin.  This argues strongly against our
having detected any
nebular component, as this should have a harder spectrum that
is well characterized by a power-law model with photon index in the range
$\sim$1--3 \citep[see][and references therein]{krh04}.

The absorbed flux of the source in the 0.5--2.0~keV range is
(6.3--6.9)$\times 10^{-15}$~erg~s$^{-1}$, where the range quoted
corresponds to that implied by the 68\% limits of $N_h$ and $kT$.  Thus,
the quoted flux range is an approximate but slightly overestimated 68\%
confidence range.  With only 110 source counts, {\tt XSPEC} is unable to
more precisely constrain the true 68\% confidence range for the flux while
simultaneously fitting for $N_h$ and $kT$. The low end of the flux range
corresponds to higher values of $N_h$ and lower values of $kT$; the high
end corresponds to the reverse. The unabsorbed 0.5--2.0~keV flux is
therefore relatively poorly constrained, ranging from $\sim 7 \times
10^{-14}$~erg~s$^{-1}$~cm$^{-2}$ for the high $kT$ end, to $\sim 2 \times
10^{-12}$~erg~s$^{-1}$~cm$^{-2}$ for the low $kT$ end. We note that the
maximum $N_h$ in this direction is $1.81 \times
10^{22}$~cm$^{-2}$, significantly lower than our upper 68\% confidence
limit \citep{dl90}.  This suggests that models
having lower values of $N_h$, and hence higher values of $kT$ and low
values of unabsorbed flux, are slightly favored.

\section{Discussion}

The dispersion measure toward the pulsar of 373~pc~cm$^{-3}$ implies a
distance of 4.0 -- 5.0~kpc \citep{cl01}.  We assume here a distance of
4.5~kpc.  Dispersion-measure distances are notoriously uncertain and an
independent distance estimate is obviously desirable.  We do note that the
\citet{tc93} distance estimate for \psr\ is 5.1~kpc, close to that
obtained with the more recent \citet{cl01} model, suggesting our
assumption of 4.5~kpc is not grossly incorrect.

Given the spectrum of the detected X-rays, the emission seems most likely
to be coming from the neutron-star surface.  Thermal emission from the
surface can either be from initial cooling, in which case X-rays are
emitted from the entire surface, or from heated polar caps, a by-product
of a higher-energy magnetospheric process \citep[see][for a
review]{krh04}.  In the former case, the X-ray energy source is unrelated
to the pulsar's spin-down.  In the latter case, the spin-down powers it.

Given the observed spectrum and flux of the X-ray source we detect, we may
ask which of these two mechanisms most likely accounts for the emission.
First, we consider the high-temperature range of parameter space, $kT
\simeq 0.2$~keV.  In this case, the unabsorbed flux, given the distance,
requires a source emitting radius of $\sim$1~km. This suggests heated
polar caps, in which case the emission could be strongly pulsed.  The
implied bolometric isotropic luminosity would be $2.5 \times
10^{32}$~erg~s$^{-1}$, or 0.16$\dot{E}$.  This is uncomfortably high for
polar-cap reheating models \citep{hm01a}.  Assuming 1.0~sr beaming, the
efficiency drops to 0.013, still implausibly high for a pulsar having
characteristic age 34~kyr \citep{hm01a}.  At the low-temperature range of
parameter space, we have $kT \simeq 0.12$~keV.  In this case, for the
observed unabsorbed flux at 4.5~kpc, an effective emitting radius of 22~km
is required, too high for a neutron star, even after correcting for the
gravitational distortion \citep{lp01}.  Thus, it seems likely on physical
grounds that even though Table~1 quotes 68\% confidence levels only, the
true spectral parameters are indeed bracketed in this range.

For example, for $kT \simeq 0.13$~keV (corresponding to $N_h \simeq 2
\times 10^{22}$~cm$^{-2}$), the observations can be accounted for if the
effective measured neutron-star radius is $\sim$13~km. In this case, the
unabsorbed bolometric luminosity would be $L_x \simeq 6 \times
10^{33}$~erg~s$^{-1}$ (corresponding to $L_x\simeq 9 \times 10^{29}$~erg~s$^{-1}$ in
the 2--10 keV band), or 4$\dot{E}$.  This, to our knowledge, would be the
first case of a radio pulsar having initial cooling emission that
has X-ray luminosity comparable to or greater than its $\dot{E}$. Given
that initial cooling is thought to be unrelated to spin-down, this is not
necessarily surprising.  More relevant is whether the effective
temperature is plausible for initial cooling.  For commonly assumed
neutron-star equations of state and modified URCA cooling with no exotica,
a temperature as high as 0.13~keV at an age of 34~kyr is reasonable if the
neutron star has accreted a $\sim 10^{-7}$~M$_{\odot}$ hydrogen envelope
\citep{yp04}.  In this case, however, because of the hydrogen envelope's
effect on the outgoing radiation, a black-body model as assumed here would
be overestimating the true effective temperature by as much as a factor of
$\sim$2 \citep[see, e.g.][]{pzs+01}.  Thus, the true effective temperature
may be much smaller than 0.13~keV, very much in line with predictions for
initial cooling of a neutron star of this age.

Even if $L_x > \dot{E}$, as seems likely in the case of \psr, $L_x$
in the 2--10~keV band is $\gapp$3 orders of magnitude smaller than
is observed for the five traditionally studied AXPs \citep[see, e.g.,
Table~2 in][]{msk+03}. Its spectral properties are also quite different
from those of the AXPs.  This is consistent with the findings for other
high-$B$ radio pulsars \citep{pkc00,gs03,msk+03,gklp04}. With X-ray
observations of five high-magnetic-field radio pulsars revealing luminosities
much smaller than those of the AXPs, it is becoming more difficult
to appeal to small scatter in the true $B$ fields relative to those
inferred from spin-down. Thus, it seems very plausible that the $B$
fields inferred from spin-down for AXPs and high-$B$ radio pulsars are
not reliable estimators of the true surface field, at least to within a
factor of $\sim$2.  Alternatively there could be a ``hidden'' parameter,
such as mass, that differentiates between the two populations.

Intriguingly, however, \psr's X-ray luminosity is comparable to that of
the recently identified transient AXP XTE~J1810$-$197 when in quiescence
\citep{ims+04,ghbb04}.  Moreover, the quiescent spectrum of
XTE~J1810$-$197 as observed in a serendipitous {\it ROSAT} observation
\citep{ghbb04} is comparable to that seen for \psr, i.e. well described by
a simple absorbed black body of temperature $kT \simeq 0.18$~keV. This
raises the interesting possibility that \psr, and other high-$B$ radio
pulsars, may one day emit transient magnetar-like emission, and conversely
that the transient AXPs might be more likely to exhibit radio pulsations.
Both these possibilities can be tested observationally.

We thank Pat Slane for providing early access to his data set.  We thank
Josh Grindlay and Jae Sub Hong for access to their {\it Chandra}
log~N/log~S relationship, and Alice Harding for useful conversations. VMK
was supported by an NSERC Discovery Grant and Steacie Fellowship
Supplement, and by the FQRNT and CIAR.

\newpage

\newpage
\begin{table}[t]
\begin{center}
\caption{Spectral Results}
\begin{tabular}{cc} \hline\hline
Parameter  & Value \\\hline
$N_h^a$ ($\times 10^{22}$~cm$^{-2}$) & 1.84$_{-0.77}^{+0.48}$    \\
$kT^a$  (keV)  & 0.145$_{-0.020}^{+0.053}$ \\
absorbed flux$^b$  ($\times 10^{-15}$~erg~s$^{-1}$~cm$^{-2}$) & 6.3--6.9\\
$\chi^2$/dof & 19.6/17 \\
unabsorbed flux$^c$ ($\times 10^{-14}$~erg~s$^{-1}$) & 7--200  \\\hline
\end{tabular}
\label{ta:spectrum}
\end{center}
$^a$Range of uncertainties indicates 68\% confidence intervals.\\
$^b$Absorbed flux in 0.5--2~keV.  Approximate 68\% confidence interval.\\
$^c$Unabsorbed flux in 0.5--2~keV.\\
\end{table}

\clearpage

\newpage

\end{document}